\documentclass[12pt,preprint]{aastex}

\setcitestyle{numbers}

\usepackage{graphicx}
\usepackage{epstopdf}

\makeatletter
\DeclareFontFamily{U}{tipa}{}
\DeclareFontShape{U}{tipa}{m}{n}{<->tipa10}{}
\newcommand{\arc@char}{{\usefont{U}{tipa}{m}{n}\symbol{62}}}%

\newcommand{\arc}[1]{\mathpalette\arc@arc{#1}}

\newcommand{\arc@arc}[2]{%
  \sbox0{$\m@th#1#2$}%
  \vbox{
    \hbox{\resizebox{\wd0}{\height}{\arc@char}}
    \nointerlineskip
    \box0
  }%
}
\makeatother

\begin{document}

\title{Golden Elliptical Orbits in Newtonian Gravitation}

\author{Dimitris M. Christodoulou}

\begin{abstract}
In spherical symmetry with radial coordinate $r$, classical Newtonian gravitation supports circular orbits and, for $-1/r$ and $r^2$ potentials only, closed elliptical orbits [1]. Various families of elliptical orbits can be thought of as arising from the action of perturbations on corresponding circular orbits. We show that one elliptical orbit in each family is singled out because its focal length is equal to the radius of the corresponding unperturbed circular orbit. The eccentricity of this special orbit is related to the famous irrational number known as the golden ratio. So inanimate Newtonian gavitation appears to exhibit (but not prefer) the golden ratio which has been previously identified mostly in settings within the animate world.

\end{abstract}

\section{Introduction}\label{intro}

In 1873, J. Bertrand [1] proved that the only spherically symmetric gravitational potentials that can support bound closed noncircular orbits are the Newton-Kepler $-1/r$ potential [12] and the isotropic Hooke $r^2$ potential [8], where $r$ is the spherical radial coordinate with respect to the central mass that generates the potential. The elliptical orbits in these two potentials were already known to I. Newton [12].

Both potentials also support circular orbits and the elliptical orbits can be thought of as arising from such circular orbits perturbed by disturbances of any arbitrary amplitude. Given a circular orbit of radius $r_o$, an infinite family of elliptical orbits can thus be obtained with eccentricities in the range $0 < e < 1$, where $e$ is related to the ratio of semiaxes $b/a$ of the ellipses by
\begin{equation}
e = \sqrt{1 - \frac{b^2}{a^2}}~ .
\label{ecc}
\end{equation}

In recent work [3], we found a new geometric property that characterizes each family of eliptical orbits and this property switches between the two potentials in a highly symmetric fashion: the circular radius $r_o$ is the harmonic mean of the radii of the turning points $r_{max}=a(1+e)$ and $r_{min}=a(1-e)$ of the ellipses in a $-1/r$ potential; whereas $r_o$ is the geometric mean of of the turning points $r_{max}=a$ and $r_{min}=b$ of the ellipses in an $r^2$ potential. 
For the reader's convenience, we summarize in Section~\ref{geo} the derivation of these properties and we proceed in Section~\ref{new} to search in each family for special orbits with additional geometric properties. Interestingly, we find that Newtonian gravitation singles out some elliptical orbits whose eccentricities are related to the golden ratio conjugate [14]
\begin{equation}
\varphi^* \equiv \frac{\sqrt{5}-1}{2} \approx 0.618\, .
\label{golden}
\end{equation}
The importance of the result lies in the fact that this famous irrational number is not introduced by the geometry of space as in the ubiquitous case of $\pi$ in Euclidean spaces; the golden ratio conjugate is instead singled out by the dynamics of the noncircular orbits. We discuss our conclusions further in Section~\ref{disc}.

\section{Known Geometric Properties of Elliptical Orbits}\label{geo}

We apply the laws of energy and angular momentum conservation to a family of elliptical orbits that arise from disturbances acting on a given circular orbit of radius $r_o$. We distinguish two cases, a Newton-Kepler potential and an isotropic Hooke potential.

\begin{figure}
\begin{center}
    \leavevmode
    \includegraphics[trim=.1cm .1cm .1cm .1cm, clip, angle=0,width=15 cm]{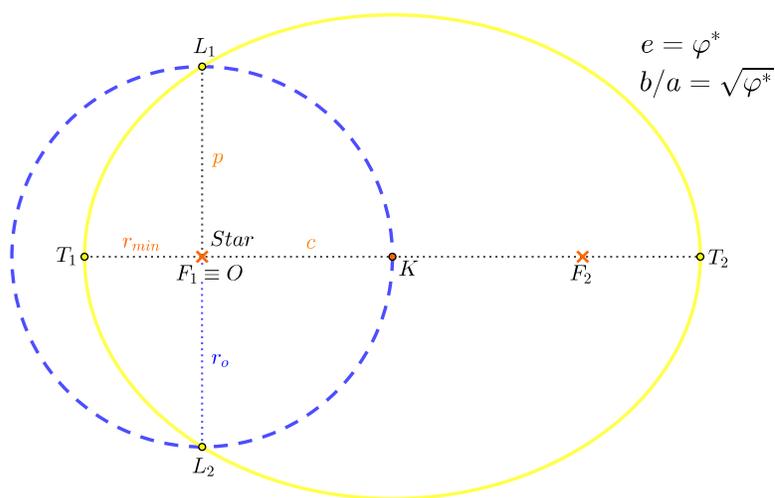}
\caption{Schematic diagram of a circular orbit $O(r_o)$ and the associated golden elliptical orbit in a Newtonian gravitational potential of the form $-1/r$ due to a star located at focus $F_1$. This focus coincides with the center $O$ of the circular orbit. Point $K$ on $\odot O$ is the center of the ellipse; $T_1$ and $T_2$ are the turning points; and $L_1$ and $L_2$ are the endpoints of the latus rectum. The golden ellipse is singled out because of the equality $c=p=r_o$ and its eccentricity $e=\phi^*\approx 0.618$.
\label{fig1}}
  \end{center}
\end{figure}

\subsection{Newton-Kepler $-1/r$ Potential}\label{newt1}

Consider an equilibrium orbit with radius $r=r_o$ in a $-1/r$ potential and assume that this orbit is perturbed to an elliptical shape with turning points $r_{min}=r_o - A_1$ and $r_{max}=r_o + A_2$, where $0<A_1<r_o$ and $A_2>A_1$ (Figure~\ref{fig1}). At the turning points $T_1$ and $T_2$ shown in Figure~\ref{fig1}, the radial velocity is zero ($dr/dt=0$) and the total energy per unit mass can then be written as [7]
\begin{equation}
{\cal E} = \frac{{\cal L}^2}{2r^2} - \frac{{\cal GM}}{r} \, ,
\label{en0}
\end{equation}
where ${\cal G}$ is the gravitatonal constant, ${\cal M}$ is the mass that generates the potential, and
the specific angular momentum satisfies ${\cal L}^2 = {\cal GM}r_o = {\rm const}.$, hence eq.~(\ref{en0}) can be written in the form
\begin{equation}
\frac{\cal E}{\cal GM} = \frac{r_o}{2r^2} - \frac{1}{r} = {\rm const.}\, , ~~ {{\rm if}~ dr/dt = 0}\, .
\label{en1}
\end{equation}

Applied to the turning points, this equation yields
\begin{equation}
\frac{1}{A_1} - \frac{1}{A_2} = \frac{2}{r_o} \, ,
\label{en3}
\end{equation}
or equivalently
\begin{equation}
\frac{1}{r_{min}} + \frac{1}{r_{max}} = \frac{2}{r_o} \, .
\label{en3b}
\end{equation}
This last equation shows that, in a $-1/r$ potential, the circular equilibrium radius $r_o$ is the harmonic mean of the radii of the turning points $r_{min}$ and $r_{max}$ of the elliptical orbits [3].

\begin{figure}
\begin{center}
    \leavevmode
    \includegraphics[trim=.1cm .1cm .1cm .1cm, clip, angle=0,width=15 cm]{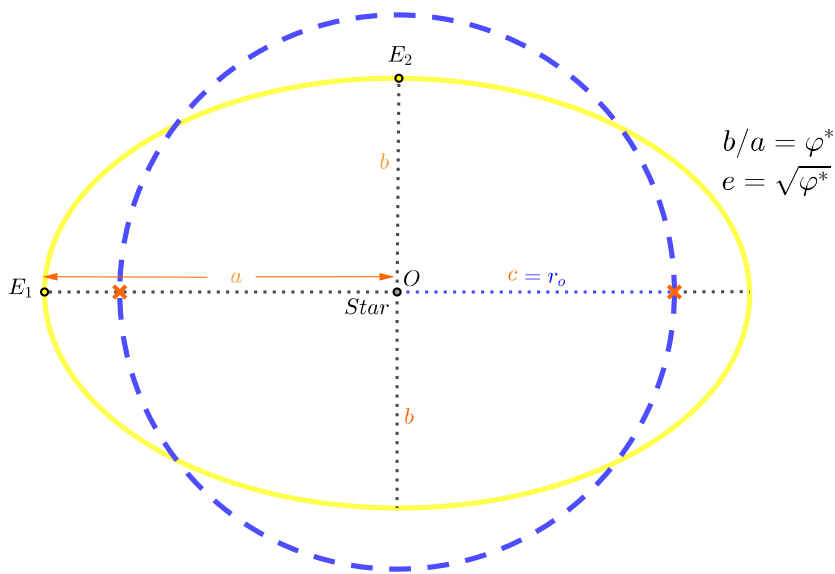}
\caption{Schematic diagram of a circular orbit $O(r_o)$ and the associated golden elliptical orbit in an isotropic Hooke gravitational potential of the form $r^2$ due to a star located at center $O$. Unlike in the Newtonian case, the gravitational force always points toward $O$ in this case. Points $E_1$ and $E_2$ are the endpoints of the semiaxes $a$ and $b$ of the ellipse. The golden ellipse is singled out because of the equality $c=r_o$, where $c$ is the focal length, and its eccentricity $e=\sqrt{\phi^*}\approx 0.786$.
\label{fig2}}
  \end{center}
\end{figure}

\subsection{Isotropic Hooke $r^2$ Potential}

Consider an equilibrium orbit with radius $r=r_o$ in a $\Omega^2 r^2/2$ potential ($\Omega=$const.) and assume that this orbit is perturbed to an elliptical shape with turning points $r_{min}=b$ and $r_{max}=a$, where $a$ and $b<a$ are the semiaxes of the ellipse (Figure~\ref{fig2}). Here the radii of the turning points take these special values because in the $r^2$ potential the gravitational force always points toward the center of the ellipse where the central mass is located.

At the turning points $E_1$ and $E_2$ shown in Figure~\ref{fig2}, the radial velocity is zero  ($dr/dt=0$) and the total energy per unit mass can then be written as
\begin{equation}
\frac{\cal E}{\Omega^2/2} = \frac{r_o^4}{r^2} + r^2 = {\rm const.}\, , ~~ {{\rm if}~ dr/dt = 0}\, ,
\label{en4}
\end{equation}
where the constant specific angular momentum is given by ${\cal L}=\Omega r_o^2$ in this case.

Applied to the turning points, this equation yields
\begin{equation}
r_{max}r_{min} = ab = r_o^2 \, .
\label{en5b}
\end{equation}
This last equation shows that, in an isotropic $r^2$ potential, the circular equilibrium radius $r_o$ is the geometric mean of the semiaxes $a$ and $b$ of the elliptical orbits [3].

\section{New Geometric Properties of Elliptical Orbits}\label{new}

The fundamental difference between the elliptical orbits in the above two cases is imposed by the dynamics: the gravitational force always points toward focus $F_1$ in Figure~\ref{fig1}, whereas it always points toward center $O$ in Figure~\ref{fig2}. This difference is responsible for producing equations~(\ref{en3b}) and~(\ref{en5b}) and it also leads to several more geometric properties of the ellipses in each case, as we describe below.

\subsection{Newtonian Ellipses}\label{newt2}

Using the well-known equations of the ellipse with semiaxes $a$ and $b$, viz.
\begin{equation}
a = (r_{max} + r_{min})/2 \, ,
\label{p1}
\end{equation}
and 
\begin{equation}
b = \sqrt{r_{max} r_{min}} \, ,
\label{p2}
\end{equation}
equation~(\ref{en3b}) takes the form
\begin{equation}
a r_o = b^2 \, ,
\label{p3}
\end{equation}
which indicates that, in Newtonian ellipses, the semiminor axis $b$ is the geometric mean of $a$ and $r_o$. More importantly, the semiaxes cannot be chosen independently in a particular family corresponding to a circular radius $r_o$. Since the semiaxes are also related to the eccentricity $e$ (eq.~(\ref{ecc})), each family of ellipses and the values of $a$ and $b$ are fully determined by the choice of $e$ for fixed $r_o$.

Written in the equivalent form
\begin{equation}
r_o = \frac{b^2}{a}\equiv p \, ,
\label{p4}
\end{equation}
equation~(\ref{p3}) indicates that $r_o$ is also equal to the semilatus rectum $p$.

Combining equations~(\ref{p4}) and~(\ref{ecc}), we find that
\begin{equation}
r_o = a (1-e^2) = b \sqrt{1-e^2} \, ,
\label{p5}
\end{equation}
and using the definition of the focal length $c$, i.e., $c=ae$, this equation takes the form
\begin{equation}
r_o = c \left(\frac{1-e^2}{e}\right) \, .
\label{p6}
\end{equation}
Equation~(\ref{p5}) shows that there are no ellipses in which one or the other semiaxis equals $r_o$. But equation~(\ref{p6}) indicates that there exists a special ellipse in the family for which the focal length $c$ equals $r_o$ (Figure~\ref{fig1}). The eccentricity of the special ellipse is found by setting $c=r_o$ in equation~(\ref{p6}), in which case we find that
\begin{equation}
e^2 + e - 1 = 0 \, ,
\label{p7}
\end{equation}
whose solution is the golden ratio conjugate shown in equation~(\ref{golden}) above.
To summarize, the only special ellipse that is singled out by the Newton-Kepler dynamics for a given $r_o$ is the golden ellipse with $e=\varphi^*$ and $c=p=r_o$, whereas all other ellipses in the family obey only $p=r_o$.

\subsection{Hookean Ellipses}

We have already seen (eq.~(\ref{en5b})) that in a family of Hookean ellipses, the radius $r_o$ of the corresponding circular orbit is the geometric mean of the semiaxes $a$ and $b$, which also implies that the areas of these ellipses are all equal to $\pi r_o^2$. Therefore, the semiaxes cannot be chosen independently within a particular family. Since the semiaxes are also related to the eccentricity $e$ (eq.~(\ref{ecc})), each family of ellipses and the values of $a$ and $b$ are fully determined by the choice of $e$ for fixed $r_o$.

Using equations~(\ref{en5b}) and~(\ref{ecc}), we find that
\begin{equation}
r_o = a (1-e^2)^{1/4} = b (1-e^2)^{-1/4}  \, ,
\label{hp5}
\end{equation}
and the semilatus rectum (eq.~(\ref{p4})) can be written as
\begin{equation}
p = r_o (1-e^2)^{3/4} \, .
\label{hp4}
\end{equation}
Hence, these ellipses are very different than the Newtonian ellipses.

Using the definition of the focal length $c=a e$, equation~(\ref{hp5}) takes the form
\begin{equation}
r_o = c \left[\frac{(1-e^2)^{1/4}}{e}\right] \, ,
\label{hp6}
\end{equation}
which indicates that there exists a special ellipse in the family for which the focal length $c$ equals $r_o$. The eccentricity of the special ellipse is found by setting $c=r_o$ in equation~(\ref{hp6}), in which case we find that
\begin{equation}
e^4 + e^2 - 1 = 0 \, ,
\label{hp7}
\end{equation}
whose solution is $e=\sqrt{\varphi^*}$. This ellipse is plotted in Figure~\ref{fig2}. Since $c=r_o$, an additional property is that its focal length is the geometric mean of its semiaxes, viz.
\begin{equation}
c^2 = ab \, .
\label{hp8}
\end{equation}

\section{Discussion}\label{disc}

The main result of this work is the appearance of the golden ratio conjugate $\varphi^*$ (eq.~(\ref{golden})) in Newtonian dynamics that predicts the existence of families of elliptical orbits only in $-1/r$ and $r^2$ spherically symmetric gravitational potentials [1]. In each case, the eccentricity $e$ of the golden elliptical orbit is related to $\varphi^*$ ($e=\varphi^*$ and $e=\sqrt{\varphi^*}$, respectively); these relations appear in the special case in which the focal length of the ellipse is equal to the radius of the corresponding circular orbit of the family ($c=r_o$ in Section~\ref{new} above).

A literature search shows that various authors use two different definitions of the golden ellipse: some authors [9, 10] define as golden the ellipse that is inscribed in a golden rectangle; others [13] construct the golden ellipse from golden right triangles in which case it is the eccentricity that is equal to the golden ratio conjugate. This latter definition is applicable to our Newtonian ellipses in which $b/a=\sqrt{\varphi^*}$. On the other hand, the former definition is applicable to our Hookean ellipses that have $b/a =\varphi^*$ and they can be inscribed in a golden rectangle.

Most of the appearances of $\varphi^*$ in the animate world are well-known if not famous. The golden ratio has been identified in phyllotaxis of plants [4]; in human constructions, including the Parthenon and the Egyptian pyramids [11]; in the honeycombs of bees [6]; and recently in the shapes of red blood cells [15].  
Lately, the golden ratio has also appeared in the inanimate world and our work concerning Newtonian dynamics falls in this category. Two more examples concern black holes [5] and cosmological theories [2], although the interested reader may easily track down more cases in solar neutrino mixing and quantum mechanics.

In the inanimate world, the golden ratio is usually introduced by the geometry of space or spacetime, but this is not the case in Newtonian dynamics. In our case, Newtonian gravitation allows for infinite families of elliptical orbits for a given circular equilibrium orbit, so the theory does not show a preference for ellipses with eccentricities $e=\varphi^*$. The golden ellipses are special only because they exhibit additional geometric properties, as we described in Sections~\ref{newt1} and~\ref{newt2}. 

In our solar system, no planetary orbit has orbital eccentricity anywhere near the golden value of 0.618; our neighboring planets and the dwarf planets all move in fairly circular orbits with $e<0.442$. Furthermore, we searched the exoplanet database (//exoplanets.org) that currently lists nearly 3,000 exoplanets with confirmed orbits and we found only three planets with $e=0.610$, two planets with $e=0.630$, and no other planets with intermediate eccentricities. So exoplanetary orbits appear to support the theoretical picture of no preference for the golden Newtonian elliptical orbits.


\vskip0.25in
\noindent
\author{Dimitris M. Christodoulou: Department of Mathematical Sciences, University of Massachusetts \\ Lowell, Lowell, MA 01854, USA} \\
{\it Email address}: {\tt dimitris\_christodoulou@uml.edu}

\end{document}